# Theory of Melting Lines with Variable Enthalpy of Fusion: Independent Verification of Phonon Theory of Liquids - Predicted Parabolic Behavior

**Anthony N. Papathanassiou**

National and Kapodistrian University of Athens, Physics Department, Condensed matter Physics Section, Panepistimiopolis, 15784 Zografos, Athens, Greece

To address the complex thermodynamics of the solid-liquid transition, we develop an analytical model by modifying the Clausius-Clapeyron (CC) equation to account for a variable enthalpy of fusion along the melting line (ML). This model draws upon theoretical and experimental progress demonstrating that the ratio of isobaric heat capacity to the thermal volume expansion coefficient in the solid state is a material constant that correlates directly with the fusion enthalpy and coexisting phase specific volumes. Differentiation of this modified CC relation yields a second-order differential equation governing the low-pressure melting curve. By imposing appropriate physical boundary conditions, physically acceptable approximate parabolic solutions are derived. Rooted in solid-state lattice anharmonicity and variable fusion enthalpy, this work provides an independent theoretical derivation for the universal parabolic behavior of low-pressure melting curves. By complementing recent formulations from the Phonon Theory of Liquids [Trachenko, Phys. Rev. E 109, 034122 (2024)], this framework helps establish a dual theoretical foundation for solid-liquid phase boundary thermodynamics.

# I. INTRODUCTION

The solid-liquid transition represents one of the most fundamental and universally recognized **phase changes** in condensed matter physics. In a pressure-temperature $(P-T)$ phase diagram, the boundary separating these two states is defined as the melting line (ML). The thermodynamic relationship governing the boundary among different states of matter is classically described by the Clausius-Clapeyron (CC) equation:

$$\frac{dP}{dT} = \frac{\delta S}{\delta v} \tag{1}$$

where $\frac{dP}{dT}$ denotes the slope of a phase coexistence curve and $\delta S$ and $\delta v$ denote the entropy and specific volume changes across the phase transition, respectively.[1, 2] In literature, it is assumed that $\delta v$ and the latent heat of transformation $(\delta H)$, representing the enthalpy change required to induce a phase transition at constant pressure and temperature, do not change significantly upon small changes in temperature or pressure. Hence, for a solid-liquid transition: $\delta S = \delta H / T_m$, where $T_m$ denotes the melting temperature.[1] For phase transitions involving a non-condensed gaseous state (namely, sublimation and boiling) the phase boundaries are well described by universal analytical equations. These classical formulations are readily derived by integrating the CC Eq. (1), facilitated by the assumption that the gas phase lacks significant cohesion energy and the latent heat is constant. However, defining an analogous universal equation for MLs has been a challenging task.[3] Because both solids and liquids are highly dense, strongly interacting condensed states, predicting the melting trajectory requires a comprehensive equation of state for liquid thermodynamics.[2,4,5]

Lacking a definitive consensus on liquid state theory, most models predicted the solid-liquid transition based exclusively on the physical and structural instabilities of the crystalline solid. The prominent early Lindemann criterion states that a solid melts when its atoms' root-mean-square displacement reaches a critical fraction of the interatomic distance.[6,7] Among various one phase theories, a critical hopping rate model appeared recently.[8] The Simon-Glatzel equation frequently models conventional melting curves that rise with pressure due to denser packing and atomic repulsion.[9,10] These heuristics estimate pressure dependence but ignore the emergent liquid's thermodynamics. Isomorph theory bridges solid and liquid-state thermodynamics.[11] For systems with strong virial-potential correlations, by proving the ML is an "isomorph curve" - where dimensionless structure and dynamics are invariant - Dyre and coworkers rationalized empirical rules like the Lindemann criterion, successfully.[11,12]

A novel theory yielding a universal analytical equation for MLs was proposed recently by Trachenko.[5] A universal, analytic two-phase theory of MLs emerged from the Phonon Theory of Liquid (PTL) Thermodynamics.[13] Balancing a liquid's vibrational and configurational entropy to determine the solid-liquid entropy difference ($\delta S$) along a melting curve remains highly challenging.[3] Hence, it is hard to determine the term $\delta S$ on the right hand side of Eq. (1). To bypass this, the PTL theory utilizes collective phonon-like excitations to quantify liquid thermodynamics. Reformulating the CC equation based on inter-phase energy differences yielded a second-order differential equation. Unlike historical models that rely on constant latent heat ($\delta H$) to derive the properties of sublimation and boiling lines[1], Trachenko's theory is fundamentally built upon the premise that $\delta H$ is variable accurately predicted that melting pressure depends on the square of the melting temperature, aligning with experimental data across diverse material classes.[5] This model is a two-phase theory and different from earlier one-phase models, which considered how the properties of solids change upon melting, and left liquids out from consideration.

The aim of the present work is the search of universal functions describing the melting curves of simple materials by modifying the CC Eq. (1), regarding a variable latent heat ($\delta H$) utilizing recent theoretical progress, according to which the heat capacity of a crystalline solid is the sum of two components: a vibrational component and a volume component.[17,18] The anharmonic component dominates at high temperatures, particularly near the melting point, and is expressed exclusively as a function of the liquid and solid specific volumes at the melting point.[19,20] Differentiating the CC equation with respect to temperature for the low-pressure regime across the ML, a simple second order differential is obtained, an approximate solution of which is a parabolic function $P(T_m)$.

The present work, together with that published recently by Trachenko[5], introduces a class of thermodynamic models for phase boundary ($P-T$) curves that bypass the historical assumption of an arbitrarily constant latent heat ($\delta H$). The idea of considering a variation of the fusion enthalpy along the melting line was employed earlier[14,15]. Particularly, Kechin found a universal equation that can appropriately describe a melting processes.[16] Even with different theoretical starting points - Trachenko's theory being rooted in the Phonon Theory of Liquids (PTL)[13,5], and the present theory in the anharmonicity on the heat capacity ($C_p$) of crystalline solids at elevated temperatures[17,18] - both approaches converge on a common approximate analytical second-degree polynomial function for $P(T_m)$ for low pressure values, namely from atmospheric pressure to marginally higher values. The present work provides an independent theoretical origin for this parabolic behavior, utilizing recent progress regarding the heat capacity of crystalline solids.

## II. RESULTS AND DISCUSSION

### A. Effect of the solid-state anharmonicity on the latent heat of fusion

In crystalline solids, the total lattice vibrational energy comprises both kinetic and potential components. The kinetic energy dictates the temperature-dependent heat capacity, whereas the potential energy arises from atomic displacements from equilibrium, coupling it directly to the macroscopic volume of the crystal. Therefore, heat capacity can be decoupled into isochoric and volume-dependent (dilatational) components.[17,18,19,20] The thermal behavior of solids at low temperatures is successfully described by the harmonic oscillator frameworks of the Einstein and Debye models. However, at elevated temperatures, anharmonicity in the lattice vibrations causes the measured isobaric heat capacity ($C_P$) to positively deviate from the classical Dulong-Petit limit of $3R$ (with $R$ representing the ideal gas constant), due to anharmonic effects. Tang *et al* [17,18,19,20] distinguished kinetic from potential contributions to the heat capacity of a crystalline solid:

$$C_P = C_T + C_{TE} \ . \tag{2}$$

The term $C_T$ ($C_T \propto f_D$, where $f_D$ is the Debye function) is the temperature dependent component and $C_{TE}$ is the volume dependent one. Within this theory, $C_{TE} = \varepsilon\beta$, where the scaling parameter $\varepsilon$ is a crystal's constant and $\beta \equiv \frac{1}{V}\left(\frac{\partial V}{\partial T}\right)_P$ is the thermal expansion coefficient. The scaling of heat capacity and thermal expansion at elevated temperatures was verified for various benchmark materials.[19] At elevated temperatures, the enthalpy ($H_{TE}$) is derived from the thermodynamic relation $C_{TE} \equiv \left(\frac{\partial H_{TE}}{\partial T}\right)_P$:

$$H_{TE} = \varepsilon ln\left(\frac{v}{v_0}\right) + H_{TE,ref} \tag{3}$$

$v$ denotes the specific volume, $v_0$ and $H_{TE,ref}$ denote reference volume and enthalpy values at temperature $T_0$. For a wide variety of solid materials, $H_{TE,ref}(T = 0K) \cong 0$.

While Eq. (3) has its origin to solid state theory of crystals, it was subsequently extended and experimentally verified to hold across the melting line: Eq. (3) Tang et al[19] validated that Eq. (3) can describe a melting process: The ratio of the isobaric heat capacity to the thermal volume expansion coefficient ($C_{TE} = \varepsilon\beta$ reads $\varepsilon = \frac{C_{P,S}}{\beta_S}$ , where $C_{P,S}$ and $\beta_S$ denote the measured heat capacity and thermal expansion coefficient) *in solid state* is a material's constant that scales the fusion enthalpy ($\delta H$) to the specific volume change ($\delta lnv$) during a transition from the solid to liquid state. Consider isobaric heating of a solid to its melting temperature, $T_m$. During the melting process, the temperature remains constant while the supplied thermal energy is absorbed as the latent heat of fusion, facilitating the solid-liquid phase transition. Associated latent heat is known as the enthalpy of fusion $\delta H$. For a melting transition, $ln\left(\frac{v_L}{v_S}\right) = lnv_L - lnv_S = \delta lnv$, where $v_L$ and $v_S$ denote the specific volumes of the liquid and the solid, respectively.[19] Thus, for a melting process, Eq. (3) reads:[19]

$$\delta H = \varepsilon ln\left(\frac{v_L}{v_S}\right) \tag{4}$$

The latent heat ($\delta H$) combines a *two-phase* term: $ln\left(\frac{v_L}{v_S}\right)$ with a *solid-state* one: $\varepsilon = \frac{C_{P,S}}{\beta_S}$ , where $C_{P,S}$ and $\beta_S$ denote the measured heat capacity and thermal expansion coefficient of the solid at elevated temperature up to the melting point. Indeed, Eq. (4) was justified experimentally for reference materials for two kinds of phase transitions: melting[19] and glass transition[20]. During the solid-liquid phase transition, the relative volume increase typically ranges from 10 to 15% for noble and molecular systems and from 10 to 20% for binary salts; in contrast, monoatomic metals and semiconductors demonstrate a smaller relative change of 1-5%, including instances of anomalous volume decrease.[21]

## B. Differential equation, mathematical solution, and physical constraints

### 1. Derivation and General Solution of the Differential Equation Governing Melting Curves.

First-order phase transitions are governed by the CC equation (Eq. (1)). For a solid-liquid boundary, the slope $\frac{dP}{dT_m}$ of the ML in a pressure-temperature ($P - T$) diagram is determined by the latent heat of fusion ($\delta H$) and the change in specific volume ($\delta v \equiv v_L - v_S$) between the coexisting phases.

$$\frac{dP}{dT_m} = \frac{\delta H}{T_m \delta v} \tag{5}$$

While sublimation and boiling curves are typically derived assuming a constant latent heat $\delta H$ in the CC equation[1], our derivation of MLs is omitting this abrupt condition. Instead, $\delta H$ is allowed to vary across the phase boundary, as defined by Eq. (4) and validated for the melting transition.[19] Substituting Eq. (4) into Eq. (5), we obtain:

$$\frac{dP}{dT_m} = \frac{\varepsilon ln\left(\frac{v_L}{v_S}\right)}{T_m \delta v} \tag{6}$$

**Evaluation of the temperature and pressure dependence of ε:** By rearranging the CC equation (Eq. 4) into Eq. (6), the derivative $\frac{dP}{dT_m}$ characterizing a phase transition is expressed as a function of the specific volumes of the two coexisting phases. To solve Eq. (6), it is essential to know how the parameter $\varepsilon$ depends on pressure and temperature across a first order phase transition from the solid to the liquid state. The temperature derivative of $\varepsilon$ of the solid phase along a ML is: $\frac{d\varepsilon}{dT} = \left(\frac{\partial \varepsilon}{\partial T}\right)_P + \left(\frac{\partial \varepsilon}{\partial P}\right)_T \frac{dP}{dT}$. By definition, $\varepsilon$ is the proportionality coefficient between $C_P$ and $\beta$ of a solid measured at various temperatures at ambient pressure. Hence, $\left(\frac{\partial \varepsilon}{\partial T}\right)_P = 0$. The value of the term $\left(\frac{\partial \varepsilon}{\partial P}\right)_T$ can be estimated in two ways: (i) As $\varepsilon \equiv \frac{C_P}{\beta}$ in the solid state, we can alternatively examine how the nominator and denominator of the fraction $\frac{C_P}{\beta}$ are varying on pressure. They both decrease on compression according to established thermodynamic expressions: $\left(\frac{\partial C_P}{\partial P}\right)_T = -T\left(\frac{\partial^2 V}{\partial T^2}\right)_P = -VT\left[\beta^2 + \left(\frac{\partial \beta}{\partial T}\right)_P\right] < 0$ and $\left(\frac{\partial \beta}{\partial P}\right)_T = -\left(\frac{\partial \kappa_T}{\partial T}\right)_P < 0$. If the numerator and denominator decrease at comparable rates with increasing pressure, then, $\varepsilon$ exhibits a weak dependence on $P$, i.e., $\left(\frac{\partial \varepsilon}{\partial P}\right)_T \approx 0$. By selecting a sufficiently narrow pressure range, $\varepsilon$ can be treated as constant. (ii) Across a solid to liquid phase transition, Eq. (4) reads: $\varepsilon = \delta H/ln\left(\frac{v_L}{v_S}\right)$.[19] We can therefore estimate how the parameter $\varepsilon$ changes along a $P(T_m)$ curve by using experimentally determined values of the fusion enthalpy ($\delta H$) and specific volumes ratio ($\frac{v_L}{v_S}$) as a function of pressure. In the low-pressure region (from ambient pressure to 50 GPa) $\delta H$ increases at a rate of 0.7%/GPa while $ln\left(\frac{v_L}{v_S}\right)$ decreases by 0.5%/GPa for body-centered cubic (bcc) tungsten.[22] For cobalt, $\delta H$ increases at a rate of 1%/GPa while $ln\left(\frac{v_L}{v_S}\right)$ decreases by 0.8%/GPa from ambient pressure to 30 GPa.[23] These examples indicate that, in the low-pressure regime, the maximum variation of $\delta H/ln\left(\frac{v_L}{v_S}\right)$ does not exceed 2% per GPa. Similarly, the parameter $\varepsilon$ varies by no more than 2%/GPa. In conclusion, $\varepsilon$ exhibits a weak pressure dependence at low pressures; hence: $\left(\frac{\partial \varepsilon}{\partial P}\right)_T \approx 0$ . Given these assumptions and constraints, $\frac{d\varepsilon}{dT_m} = \left(\frac{\partial \varepsilon}{\partial T}\right)_P + \left(\frac{\partial \varepsilon}{\partial P}\right)_T \frac{dP}{dT} \approx 0$. This simplification facilitates the solution to Eq. (6), enabling the derivation of an analytical expression, $P(T_m)$, for the melting curve.

Differentiating Eq. (6) with respect to temperature, we get:

$$(v_L - v_S)\frac{dP}{dT_m} + T_m \frac{d(v_L - v_S)}{dT_m}\frac{dP}{dT_m} + T_m(v_L - v_S)\frac{d^2P}{dT_m{}^2} = \varepsilon\left(\frac{dlnv_L}{dT_m} - \frac{dlnv_S}{dT_m}\right) \tag{7}$$

$$T_m(v_L - v_S)\frac{d^2P}{dT_m{}^2} + (v_L - v_S)\left[1 + T_m \frac{d(v_L - v_S)}{dT_m}\frac{dP}{dT_m}\right] = \varepsilon\left(\frac{dlnv_L}{dT_m} - \frac{dlnv_S}{dT_m}\right) \tag{8}$$

To express the derivatives $\frac{dlnv_L}{dT_m}$ and $\frac{dlnv_S}{dT_m}$ in terms of thermophysical properties of the liquid and solid states, we use the thermodynamic identity $\frac{dV}{dT} = \left(\frac{\partial V}{\partial T}\right)_P + \left(\frac{\partial V}{\partial P}\right)_T \frac{dP}{dT} = \beta V - \frac{V}{B}\frac{dP}{dT}$, which gives $\frac{dlnV}{dT_m} = \beta - \frac{1}{B}\frac{dP}{dT_m}$, where $\beta$ is thermal expansion coefficient and $B$ is the isothermal bulk modulus. Substituting this relation into Eq. (8)and dividing by $T_m(v_L - v_S)$ yields:

$$\frac{d^2P}{dT_m{}^2} + \left\{\frac{1}{T_m} + \frac{dln(v_L - v_S)}{dT_m} + \frac{\varepsilon}{T_m(v_L - v_S)}\left(\frac{1}{B_L} - \frac{1}{B_S}\right)\right\}\frac{dP}{dT_m} = \frac{(\beta_L - \beta_S)}{T_m(v_L - v_S)}\varepsilon \tag{9}$$

Eq. (9) is written in a simple form:

$$\frac{d^2P}{dT_m{}^2} + b\frac{dP}{dT_m} = d \tag{10}$$

where the parameters $b$ and $d$ are defined as:

$$b(T_m) \equiv \frac{1}{T_m} + \frac{dln(v_L - v_S)}{dT_m} + \frac{\varepsilon}{T_m(v_L - v_S)}\left(\frac{1}{B_L} - \frac{1}{B_S}\right) \quad (11)$$

$$d(T_m) \equiv \frac{(\beta_L - \beta_S)}{T_m(v_L - v_S)}\varepsilon \quad (12)$$

**Signs and relative magnitude of $\boldsymbol{b(T_m)}$ and $\boldsymbol{d(T_m)}$:** The melting transition is usually accompanied by a swelling of the molar volume ($\delta v \equiv v_L - v_S > 0$). A material exhibits higher isothermal compressibility (or lower isothermal bulk modulus) and thermal expansion coefficient in its liquid state compared with those of its solid state. In normal cases, $\beta_L \gg \beta_S$ and $B_S > B_L$. For example, typical ranges of $\beta_L$ and $\beta_S$ are $10^{-5}$ -$10^{-3}$ $K^{-1}$ and $10^{-6}$ - $10^{-5}$ $K^{-1}$, respectively, whilst typical ranges of $B_S$ and $B_L$ are 30 - 450 GPa and 1 - 3 GPa, respectively. So, $\left(\frac{1}{B_L} - \frac{1}{B_S}\right) > 0$ and $\beta_L - \beta_S > 0$. Besides, $\delta v$ remains nearly constant along the ML: for example, the ratio $\frac{v_L}{v_S}$ remains constant along the ML for the inverse power potentials[24], while measurements of $\delta v$ of indium along its ML[25], yielded: $\frac{dln(v_L - v_S)}{dT_m} \cong -0.002\ K^{-1}$, which is small (compared to the value of the term $\frac{\varepsilon}{(v_L - v_S)}\left(\frac{1}{B_L} - \frac{1}{B_S}\right)$). As a result, the sign of both coefficients $b$ and $d$ are strictly positive: $b(T_m) > 0$ and $d(T_m) > 0$.

**Constancy of the ratio $\frac{\boldsymbol{d}(T_m)}{\boldsymbol{b}(T_m)}$ and relation $\boldsymbol{d}(T_m) \gg \boldsymbol{b}(T_m)$:** Although both $\boldsymbol{b}(T_m)$ and $\boldsymbol{d}(T_m)$ explicit depend on $T_m$ through the denominator factor $T_m(v_L - v_S)$, their ratio $\boldsymbol{d}(T_m)/\boldsymbol{b}(T_m)$ exhibits a fundamental property: because the term containing bulk moduli dominates the expression for $\boldsymbol{b}(T_m)$, the factor $T_m(v_L - v_S)$ cancels out exactly: $\frac{\boldsymbol{d}(T_m)}{\boldsymbol{b}(T_m)} \cong \frac{(\beta_L - \beta_S)}{\left(\frac{1}{B_L} - \frac{1}{B_S}\right)}$: As discussed above, $\beta_L \gg \beta_S$ and $B_S > B_L$, hence: $\frac{\boldsymbol{d}(T_m)}{\boldsymbol{b}(T_m)} \approx \beta_L B_L$. The latter implies that $\frac{\boldsymbol{d}(T_m)}{\boldsymbol{b}(T_m)} \approx C$, where $C$ denotes a constant, if pressure varies within a narrow range. Besides, the value of the product $\beta_L B_L$ typically ranges from $10^4$ to $3 \times 10^6$ Pa/K[5] implying that $\boldsymbol{d}(T_m) \gg \boldsymbol{b}(T_m)$.

Since $\frac{\boldsymbol{d}(T_m)}{\boldsymbol{b}(T_m)} \approx C$, Eq.(10) reads: $\frac{d^2P}{dT_m{}^2} + \boldsymbol{b}(T_m)\frac{dP}{dT_m} = \boldsymbol{d}(T_m) \approx C\boldsymbol{b}(T_m)$ . Setting $\boldsymbol{u}(T_m) \equiv \frac{dP}{dT_m}$, this becomes:

$$\frac{du}{dT_m} + \boldsymbol{b}(T_m)\boldsymbol{u} = C\boldsymbol{b}(T_m) \quad (13)$$

Integrating with an integrating factor $e^{\int b(T_m)dT_m}$ yields: $\frac{dP}{dT_m}$,=C+$c_1 e^{-\int b(T_m)dT_m}$, where $c_1$ denotes an integration constant. Integrating once more with respect to $T_m$ gives the exact general solution for the melting line:

$$P(T_m) = \frac{d}{b}T_m + c_1 \int e^{-\int b(T_m)dT_m}\, dT_m + c_2 \quad (14)$$

where $c_2$ is an integration constant.

When $b(T_m)$ and $d(T_m)$ are evaluated over a narrow low-pressure temperature range where they vary slowly around average values $b_0$ and $d_0$, respectively, Eq. (14) simplifies to:

$$P(T_m) \cong \frac{d_0}{b_0}T_m - \frac{c_1}{b_0}e^{-b_0 T_m} + c_2 \quad (15)$$

Eq. (15) represents a mathematical solution to Eq. (10). However, to render the function describing a ML physically sound in the low-pressure region, precise constraints must be imposed on the constants of integration. These constraints should guarantee that the curve $P(T_m)$ remains strictly monotonically increasing and convex for all low-pressure values $P \geq 0$ and $T \geq T_0$, where $T_0$ denotes the melting temperature at zero-pressure (i.e., $P(T_0) \equiv 0$).

**Convex profile of $P(T_m)$:** Firstly, from Eq. (15), we get: $\frac{d^2P}{dT_m{}^2} = -c_1 b_0 e^{-b_0(T_m)T_m}$. Since $b_0 > 0$ and $e^{-b_0 T_m}$>0, the sign of $\frac{d^2P}{dT_m{}^2}$ is ruled entirely by the sign of $c_1$. For a physically sound melting curve that increases monotonically with pressure starting from $P(T_0) \equiv 0$ at zero pressure with an slope $\frac{\boldsymbol{d}(T_m)}{\boldsymbol{b}(T_m)} \approx \beta_L B_L$>0, boundary condition fitting requires $c_1 < 0$. Consequently $\frac{d^2P}{dT_m{}^2} = -c_1 b(T_m) e^{-b(T_m)T_m}$ is positive throughout the entire low-pressure domain. This implies that the ML is strictly convex and concaves upwards.

**Monotonicity of $P(T_m)$: D**ifferentiating Eq. (15) yields: $\frac{dP}{dT_m} = \frac{d_0}{b_0} + c_1 e^{-b_0 T_m}$.Investigation of the monotonicity of the function $P(T_m)$ requires determination of the integration constant $c_1$, which was found to be negative above ($c_1 < 0$): The melting curve is anchored by two invariant physical state points: zero-pressure melting point $(T_0, 0)$ and the triple point $(T_c, P_c)$. $(T_c, P_c)$ represents a distinct finite-pressure reference state with $T_c > T_0$ and $P_c > 0$. The first one is experimentally accessible for any material and can be predicted theoretically; the latter is thermodynamically invariant unique signature of matter. Accordingly, the integration constants are obtained:

$$c_1 = \frac{b_0[P_c - \frac{d_0}{b_0}(T_c - T_0)]}{e^{-b_0 T_0} - e^{-b_0 T_c}} \tag{16}$$

$$c_2 = \left(\frac{P_c - \frac{d_0}{b_0}(T_c - T_0)}{e^{-b_0 T_0} - e^{-b_0 T_c}}\right) e^{-b_0 T_0} - \frac{d_0}{b_0} T_0 \tag{17}$$

**Estimation of the slope of a ML:** Substituting the coefficient $c_1$ provided by Eq. (16) into $\frac{dP}{dT_m} = \frac{d_0}{b_0} + c_1 e^{-b_0 T_m}$. we get: $\frac{\mathrm{d}P}{\mathrm{d}T_m} = \frac{d_0}{b_0} + \frac{b_0\left[P_c - \left(\frac{d_0}{b_0}\right)(T_c - T_0)\right]}{e^{-b_0 T_0} - e^{-b_0 T_c}} e^{-b_0 T_m}$. Factoring $e^{-b_0 T_0}$ out of the denominator and applying the first-order Taylor expansion $e^{-x} \approx 1 - x$ with $x = b_0(T_c - T_0) \ll 1$ for a low-pressure narrow temperature range, gives: $e^{-b_0 T_0} - e^{-b_0 T_c} = e^{-b_0 T_0}\left[1 - e^{-b_0(T_c - T_0)}\right] \approx e^{-b_0 T_0}\, b_0(T_c - T_0)$. Replacing the denominator $e^{-b_0 T_0} - e^{-b_0 T_c}$ with $e^{-b_0 T_0}\, b_0(T_c - T_0)$ yields: $\frac{\mathrm{d}P}{\mathrm{d}T_m} \approx \frac{d_0}{b_0} + \frac{b_0\left[P_c - \left(\frac{d_0}{b_0}\right)(T_c - T_0)\right]}{e^{-b_0 T_0}\, b_0(T_c - T_0)} e^{-b_0 T_m}$. Canceling $b_0$ and $(T_c - T_0)$, and combining $e^{-b_0 T_m}/e^{-b_0 T_0} = e^{-b_0(T_m - T_0)}$, gives: $\frac{\mathrm{d}P}{\mathrm{d}T_m} \approx \frac{d_0}{b_0} + \left(\frac{P_c}{T_c - T_0} - \frac{d_0}{b_0}\right) e^{-b_0(T_m - T_0)}$. This shows directly that the slope transitions from $\approx \frac{P_c}{T_c - T_0}$ near $T_0$ toward the asymptotic plateau $\frac{d_0}{b_0} \approx \beta_L B_L$ as $b_0(T_m - T_0)$ increases.

Notably, a slope equal to approximately $\beta_L B_L$ was also predicted by Trachenko, whose model of MLs is derived from the Phonon Theory of Liquids (PTL).[5] The experimental justification of the common prediction of the two independent theories that the slopes of the melting curves are approximately equal $\beta_L B_L$ for various materials (such as Ar , He, $H_2$, $H_20$, In, Fe) was discussed by Trachenko.[5]

**Derivation of the universal parabolic melting law for $\boldsymbol{b_0(T_m - T_0) \ll 1}$:** By writing $e^{-b_0 T_m} = e^{-b_0 T_0} e^{-b_0(T_m - T_0)}$ and defining the rescaled integration constant $c_1^* \equiv c_1 e^{-b_0 T_0}$ (which retains the exact same sign as $c_1$ since $e^{-b_0 T_0} > 0$), Eq. (15) reads:

$$P(T_m) \cong \frac{d_0}{b_0} T_m - \frac{c_1^*}{b_0} e^{-b_0(T_m - T_0)} + c_2 \tag{18}$$

Over a narrow low-pressure interval $T_0 \le T_m \le T_c$ where $b_0(T_m - T_0) \ll 1$, the Taylor expansion of the term $e^{-b_0(T_m-T_0)}$ to second order in $\Delta T \equiv T_m - T_0$ as:

$$e^{-b_0(T_m-T_0)} \approx 1 - b_0(T_m - T_0) + \frac{1}{2}b_0^2(T_m - T_0)^2 \tag{19}$$

Substituting Eq. (19) into $Eq.(16)$ yields a second-degree polynomial in $(T_m - T_0)$:

$$P_m(T_m) \approx -\frac{c_1^* b_0}{2}(T_m - T_0)^2 + \left(\frac{d_0}{b_0} + c_1^*\right)(T_m - T_0) + \left(c_2 - \frac{c_1^*}{b_0} + \frac{d_0}{b_0}T_0\right) \tag{20}$$

Or, expanded in powers of $T_m$:

$$P_m(T_m) \approx -\frac{c_1^* b_0}{2}T_m^2 + \left(\frac{d_0}{b_0} + c_1^* + c_1^* b_0 T_0\right)T_m + \left(c_2 - \frac{c_1^*}{b_0} - \frac{c_1^* b_0}{2}T_0^2\right) \tag{21}$$

Imposing $P_m(T_0) = 0$ immediately fixes $c_2$: $c_2 = \frac{c_1^*}{b_0} - \frac{d_0}{b_0}T_0$. Imposing $P_m(T_c) = P_c$ gives $c_1^*(T_c - T_0)\left[1 - \frac{b_0}{2}(T_c - T_0)\right] = P_c - \frac{d_0}{b_0}(T_c - T_0)$, determining both integration constants uniquely:

$$c_1^* = \frac{P_c - \frac{d_0}{b_0}(T_c - T_0)}{(T_c - T_0)\left[1 - \frac{b_0}{2}(T_c - T_0)\right]} \tag{22}$$

$$c_2 = \frac{P_c - \frac{d_0}{b_0}(T_c - T_0)}{b_0(T_c - T_0)\left[1 - \frac{b_0}{2}(T_c - T_0)\right]} - \frac{d_0}{b_0}T_0 \tag{23}$$

Substituting $c_1^*$ and $c_2$ (Eqs. (22) and (23)) into Eq.(20) gives:

$$P_m(T_m) \approx \left[\frac{\frac{P_c}{T_c - T_0} - \frac{1}{2}d_0(T_c - T_0)}{1 - \frac{1}{2}b_0(T_c - T_0)}\right](T_m - T_0) + \frac{1}{2}\left[d_0 - b_0\left(\frac{\frac{P_c}{T_c - T_0} - \frac{1}{2}d_0(T_c - T_0)}{1 - \frac{1}{2}b_0(T_c - T_0)}\right)\right](T_m - T_0)^2 \tag{24}$$

The initial slope is: $\left.\frac{dP_m}{dT_m}\right|_{T_0} = \frac{d_0}{b_0} + c_1^*$.

**Concavity of the $P_m(T_m)$ parabola:** The curvature of the parabola described by Eq. (24) is: $\frac{d^2P_m}{dT_m^2} = -c_1^* b_0 = d_0 - b_0\left(\frac{\frac{P_c}{T_c - T_0} - \frac{1}{2}d_0(T_c - T_0)}{1 - \frac{1}{2}b_0(T_c - T_0)}\right) = \frac{d_0(T_c - T_0) - b_0 P_c}{(T_c - T_0)\left[1 - \frac{b_0}{2}(T_c - T_0)\right]}$. Because $b_0 > 0$ and $T_c > T_0$, strict upward concavity ($c_1^* < 0$ and $\frac{d^2P_m}{dT_m^2} > 0$ throughout $T_0 \le T_m \le T_c$) holds whenever two physical conditions are valid: (i) the denominator $1 - \frac{b_0}{2}(T_c - T_0)$ remains strictly positive across any low-pressure reference interval satisfying $T_c - T_0 < 2/b_0$. (ii) For a convex melting curve whose slope increases monotonically toward its asymptotic plateau $\frac{d_0}{b_0} \approx \beta_L B_L$, the average secant slope $\frac{P_c}{T_c - T_0}$ over $[T_0, T_c]$ is strictly less than $\frac{d_0}{b_0}$ (($\frac{P_c}{T_c - T_0} < \frac{d_0}{b_0} \approx \beta_L B_L$)), making the numerator $d_0(T_c - T_0) - b_0 P_c$ strictly positive.

## 3. Parabolic ML $P_m(T_m)$ with Variable Coefficients $b(T_m)$ and $d(T_m)$ Linearly Dependent on temperature.

Expanding $b(T_m)$ and $d(T_m)$, defined in Eqs.(11) and (12), in a Taylor series around $T_0$ up to first order in $\Delta T = T_m - T_0$ gives:

$$b(T_m) \cong b_0 + b_1(T_m - T_0) \tag{25}$$

$$d(T_m) \cong d_0 + d_1(T_m - T_0) \tag{26}$$

Evaluating Eqs. (11) and (12) and their first temperature derivatives at $T_m = T_0$ yields the zeroth-order and general first-order Taylor coefficients:

$$b_0 = \frac{1}{T_0} + \left(\frac{d \ln \delta v}{dT_m}\right)_0 + \frac{\varepsilon}{T_0\, \delta v_0}\left(\frac{1}{B_{L,0}} - \frac{1}{B_{S,0}}\right) \tag{27}$$

$$d_0 = \frac{\beta_{L,0} - \beta_{S,0}}{T_0\, \delta v_0}\, \varepsilon \tag{28}$$

$$b_1 \equiv \left(\frac{db}{dT_m}\right)_0 = -\frac{1}{T_0^2} + \left(\frac{d^2 \ln \delta v}{dT_m^2}\right)_0 + \frac{\varepsilon}{T_0\, \delta v_0}\left\{\left[\frac{d(1/B_L - 1/B_S)}{dT_m}\right]_0 - \left(\frac{1}{B_{L,0}} - \frac{1}{B_{S,0}}\right)\left[\frac{1}{T_0} + \left(\frac{d \ln \delta v}{dT_m}\right)_0\right]\right\} \tag{29}$$

$$d_1 \equiv \left(\frac{dd}{dT_m}\right)_0 = \frac{\varepsilon}{T_0\, \delta v_0}\left\{\left[\frac{d(\beta_L - \beta_S)}{dT_m}\right]_0 - \left(\beta_{L,0} - \beta_{S,0}\right)\left[\frac{1}{T_0} + \left(\frac{d \ln \delta v}{dT_m}\right)_0\right]\right\} \tag{30}$$

When the thermophysical parameters $\delta v$, $\beta_{L,S}$, and $B_{L,S}$ vary slowly in the immediate vicinity of $T_0$, the explicit $1/T_m$ dependence dominates the temperature derivatives, reducing Eqs. (29)–(30) to the compact scaling relations:

$$b_1 \approx -\frac{b_0}{T_0} \tag{31}$$

$$d_1 \approx -\frac{d_0}{T_0} \tag{32}$$

Substituting Eqs. (31) and (32) into Eqs. (25) and (26), respectively, expresses both first-order coefficient expansions solely in terms of the zero-pressure reference values $b_0$, $d_0$, and $T_0$:

$$b(T_m) \cong b_0\left(1 - \frac{T_m - T_0}{T_0}\right) \tag{33}$$

$$d(T_m) \cong d_0\left(1 - \frac{T_m - T_0}{T_0}\right) \tag{34}$$

**Small-parameter regime ($|b_0(T_m - T_0)| \ll 1$ and $|(T_m - T_0)/T_0| \ll 1$):** Denoting the melting slope by $u(\Delta T) \equiv dP_m/dT_m$ and substituting the scaling-reduced expansions of Eq. (33) into Eq. (13) yields the first-order linear differential equation:

$$\frac{du}{d(\Delta T)} + b_0\left(1 - \frac{\Delta T}{T_0}\right)u = d_0\left(1 - \frac{\Delta T}{T_0}\right) \tag{35}$$

Including the linear $-(b_0/T_0)\Delta T$ term modifies the integrating factor from the pure exponential $\exp(b_0 T_m)$ of the zeroth-order treatment to a Gaussian-modulated exponential:

$$\mu(\Delta T) = \exp\left[\int_0^{\Delta T} b_0\left(1 - \frac{\tau}{T_0}\right) d\tau\right] = \exp\left(b_0 \Delta T - \frac{b_0}{2T_0}\Delta T^2\right)$$

Multiplying Eq. (35) by $\mu(\Delta T)$ and integrating from 0 to $\Delta T$ produces the exact slope representation:

$$\frac{dP_m}{dT_m} = \exp\left(-b_0\Delta T + \frac{b_0}{2T_0}\Delta T^2\right) \cdot \left[\tilde{c}_1 + d_0 \int_0^{\Delta T} \left(1 - \frac{\tau}{T_0}\right) \exp\left(b_0 \tau - \frac{b_0}{2T_0}\tau^2\right) d\tau\right]$$

where $\tilde{c}_1 \equiv (dP_m/dT_m)_0$ is the initial slope of the melting curve at $T_0$. In the small-parameter regime ($|b_0 \Delta T| \ll 1$ and $|\Delta T/T_0| \ll 1$), expanding the exponential prefactor and integral up to second order in $\Delta T$ gives:

$$\exp\left(-b_0\Delta T + \frac{b_0}{2T_0}\Delta T^2\right) \approx 1 - b_0 \Delta T + \frac{b_0}{2}\left(b_0 + \frac{1}{T_0}\right)\Delta T^2 + \mathcal{O}(\Delta T^3) \tag{36}$$

$$d_0 \int_0^{\Delta T}\left(1 - \frac{\tau}{T_0}\right)(1 + b_0 \tau + \cdots)\, d\tau \approx d_0 \Delta T + \frac{d_0}{2}\left(b_0 - \frac{1}{T_0}\right)\Delta T^2 + \mathcal{O}(\Delta T^3) \tag{37}$$

Multiplying Eqs. (36) and (37) and retaining terms up to $\mathcal{O}(\Delta T^2)$ yields the power-series expansion of the melting slope, in which $(d_0 - b_0 c_1)$ factors out identically from the quadratic term:

$$\frac{dP_m}{dT_m} \approx \tilde{c}_1 + (d_0 - b_0 c_1)\Delta T - \frac{1}{2}\left(b_0 + \frac{1}{T_0}\right)(d_0 - b_0 \tilde{c}_1)\Delta T^2 \tag{38}$$

Integrating Eq. (38) term-by-term with respect to $\Delta T = T_m - T_0$ generates the melting pressure series up to $\mathcal{O}(\Delta T^3)$:

$$P_m(T_m) \approx \tilde{c}_2 + \tilde{c}_1(T_m - T_0) + \frac{1}{2}(d_0 - b_0 c_1)(T_m - T_0)^2 - \frac{1}{6}\left(b_0 + \frac{1}{T_0}\right)(d_0 - b_0\tilde{c}_1)(T_m - T_0)^3 \quad (39)$$

**Boundary Conditions and the Parabolic Melting Law $P_m(T_m)$:** The integration constants $\tilde{c}_1$ and $\tilde{c}_2$ are fixed uniquely by anchoring the melting curve at two thermodynamically reference states across the low-pressure interval $\Delta T_{ref} \equiv T_{ref} - T_0$, where $T_{ref} \equiv T_c(1+\lambda)$ **with** $1 \gg \lambda > 0$. A thermodynamically invariant state $(T_0, 0)$ and a general low-pressure coexistence reference point $(T_{\mathrm{ref}}, P_{\mathrm{ref}})$. the later is any experimentally measured coexistence point on the melting curve within the low-pressure validity window. Imposing $P_m(T_0) = 0$ in Eq. (39) immediately gives $\tilde{c}_2 = 0$. Imposing $P_m(T_{ref}) = P_{ref}$ in Eq. (39) incorporates the first-order $1/T_m$ scaling corrections directly into the renormalized initial slope $\tilde{c}_1$:

$$\tilde{c}_1 = \frac{\frac{P_{ref}}{\Delta T_{ref}} - \frac{1}{2}d_0\Delta T_{ref} + \frac{1}{6}d_0\left(b_0 + \frac{1}{T_0}\right)\Delta T_{ref}^2}{1 - \frac{1}{2}b_0\Delta T_{ref} + \frac{1}{6}b_0\left(b_0 + \frac{1}{T_0}\right)\Delta T_{ref}^2} \quad (40)$$

The denominator is non-zero as it has a strictly negative discriminator with respect to $\Delta T_{ref}$ (i.e., $-\frac{b_0(5b_0T_0+8)}{12T_0} < 0$). Truncating Eq. (39) at quadratic order while retaining the $1/T_m$-renormalized initial slope $\tilde{c}_1$ from Eq. (40) yields the canonical parabolic melting curve:

$$P_m(T_m) \cong \tilde{c}_1(T_m - T_0) + \frac{1}{2}(d_0 - b_0\tilde{c}_1)(T_m - T_0)^2 \quad (41)$$

Expanding $(T_m - T_0)^2 = T_m^2 - 2T_0T_m + T_0^2$ expresses the melting line in standard second-degree polynomial form:

$$P_m(T_m) \cong A_0 + A_1T_m + A_2T_m^2 \quad (42)$$

where: $A_0 = -\tilde{c}_1T_0 + \frac{1}{2}(d_0 - b_0\tilde{c}_1)T_0^2 = -\tilde{c}_1T_0 + A_2T_0^2$, $A_1 = \tilde{c}_1 - 2A_2T_0$ and $A_2 = \frac{1}{2}(d_0 - b_0\tilde{c}_1)$. From Eq. (42), the second derivative of the melting curve at $T_0$ is $d^2P_m/dT_m^2 = 2A_2 = d_0 - b_0\tilde{c}_1$. Upward concavity ($A_2 > 0$) holds provided that $\tilde{c}_1 < d_0/b_0 \approx \beta_{L,0}B_{L,0}$, where $\tilde{c}_1$ is evaluated via Eq. (40). Under the slow-variation scaling relations $b_1 \approx -b_0/T_0$ and $d_1 \approx -d_0/T_0$, the curvature factor $(d_0 - b_0\tilde{c}_1)$ cancels identically between the leading cubic and quadratic terms of Eq. (39).

## III. CONCLUSIONS

We have presented a rigorous analytical framework for the solid-liquid phase boundary that overcomes the traditional constant-latent-heat approximations of CC formulations. An analytically solvable model for the MLs was derived for the low-pressure regime by accounting for the anharmonic, volume-dependent nature of isobaric heat capacity of the solid state,enthalpy of fusion and specific volumes of coexisting phases. Subject to distinct physical constraints, the resulting expressions reduce to approximate parabolic functions describing the low pressure region of a ML. This model relies on no empirical fitting; its parameters are entirely governed by fundamental, measurable quantities, such as the bulk moduli, thermal expansion coefficients, and specific volumes of the coexisting phases, as well as the isobaric heat capacity of the solid state. The predictions of this anharmonicity-based study regarding (i) the parabolic nature of the melting curves and (ii) the magnitude of $\frac{dP}{dT_m} \approx \beta_L B_L$ are in striking agreement with those of Trachenko's theory based on the Phonon Theory of Liquids (PTL)[5]. This striking agreement, arising from completely distinct theoretical starting points, establishes the universal parabolic nature of low-pressure melting curves and established a robust dual theoretical foundation for the universal parabolic thermodynamics of solid-liquid phase transitions. Showing that two distinct theoretical entry points - liquid-state phonon theory (PTL) and solid-state lattice anharmonicity - independently converge on identical parabolic scaling laws and slopes $\frac{\mathrm{d}P}{\mathrm{d}T_m} \approx \beta_L B_L$ establishes a robust dual theoretical foundation for solid-liquid phase boundary thermodynamics.

---

[1] Landau, L. D.; Lifshitz, E. M. *Statistical Physics, Part 1*; Course of Theoretical Physics, Vol. 5; Pergamon Press: Oxford, 1970. DOI: 10.1016/B978-0-08-023039-9.X5001-8

[2] de With, G. Melting Is Well-Known, but Is It Also Well-Understood? *Chem. Rev.* **2023**, *123* (25), 13713-13795. DOI: 10.1021/acs.chemrev.3c00489

[3] Ubbelohde, A. R. *The Molten State of Matter: Melting and Crystal Structure*; John Wiley & Sons, 1978.

[4] Anderson, O. L. *Equations of State of Solids for Geophysics and Ceramic Science*; Oxford University Press: New York, 1995. DOI: 10.1093/oso/9780195056068.001.0001.

[23] Zhang, W.-J.; Peng, Y.-F.; Liu, Z.-L. Molecular Dynamics Study of Melting Curve, Entropy of Fusion and Solid–Liquid Interfacial Energy of Cobalt Under Pressure. Physica B **2014**, 440, 33–40. DOI: 10.1016/j.physb.2014.01.015.

[24] Stishov, S. M. Thermodynamics of Melting of Simple Substances. *Sov. Phys. Usp.* **1974**, *17* (6), 625–643

[25] Höhne, G. W. H.; Dollhopf, W.; Blankenhorn, K.; Mayr, P. U. On the pressure dependence of the heat of fusion and melting temperature of indium. *Thermochim. Acta* **1996**, *273*, 17–24. DOI: 10.1016/0040-6031(95)02686-X.